\documentclass{elsart5}

\usepackage{graphicx}

\usepackage{amssymb}

\begin{document}

\begin{frontmatter}

\title{The electronic structure and the phases of BaVS$_3$}

\author{P. Fazekas\corauthref{cor1}}
\ead{pf@szfki.hu}
\author{K. Penc}
\author{K. Radn\'oczi}
\corauth[cor1]{}
\address{Research Institute for Solid State Physics and Optics, Budapest 114, POB 49, H-1525 Hungary}
\author{N. Bari\v{s}i\'c}
\author{H. Berger}
\author{L. Forr\'o}
\author{S. Mitrovi\'c\thanksref{label3}}
\thanks[label3]{Present address: Department of Physics, California Institute of Technology, Pasadena, CA 91 125, USA}
\address{Institut de Physique de la Mati\'ere Complexe, EPFL, CH-1015, Lausanne, Switzerland}
\author{A. Gauzzi}
\address{Institut de Mineralogie et de Physique des Milieux Condensee, Universite de Paris 6 "Pierre et Marie Curie", F-75015 Paris France}
\author{L. Demk\'o}
\author{I. K\'ezsm\'arki}
\author{G. Mih\'aly}
\address{Institute of Physics, Budapest University of Technology and Economics, Budafoki u. 8, Budapest, H-1111 Hungary}
\received{}
\revised{}
\accepted{}


\begin{abstract}
 BaVS$_3$ is a moderately correlated $d$-electron system with a rich phase diagram. To construct the corresponding minimal electronic model, one has to decide which $d$-states are occupied, and to which extent. The ARPES experiment presented here shows that the behavior of BaVS$_3$ is governed by the coexistence of wide-band ($A_{1g}$) and narrow-band ( twofold degenerate $E$) $d$-electrons.
We sketch a lattice fermion model which may serve as a minimal model of BaVS$_3$. This serves foremost for the understanding of the metal-insulator in pure BaVS$_3$ and its absence in some related compounds. The nature of the low temperature magnetic order differs for several systems which may be described in terms of the same
electron model. We describe several recent experiments which give information about magnetic order at high pressures. In particular, we
discuss field-induced insulator-to-metal transition at slightly subcritical pressures, and an evidence for magnetic order in the high-pressure metallic phase. The phase diagram of Sr-doped BaVS$_3$ is also discussed. The complexity of the phases of BaVS$_3$ arises from the fact that it is simultaneously unstable against several kinds of instabilities.
\end{abstract}


\end{frontmatter}

\section{Introduction}\label{intr}


$3d$ transition metal (TM) compounds are expected to obey Mott physics: the relatively compact $d$-shells are associated with a $U\sim W$ (bandwidth), thus many TM oxides, halides, sulphides, etc are strongly correlated $d$-metals, or Mott insulators with localized $d$-electrons, or show transition between such states at accessible temperatures ($T$) and pressures ($p$) \cite{Imada}. Just in the borderline cases, the emerging low-$T$  order (if any) shows great variability: while in the large-$U$ limit, we find usually long-range order with a strictly local (single-$d$-shell) order parameter (OP), at $U\sim W$ the relevant spatial unit often extends over two lattice sites (dimer), or even four (tetramer, or plaquette, depending on the lattice geometry).

Stoichiometric good-quality BaVS$_3$ is a moderately correlated $3d$ sulphide with a metal--insulator transition and a variety of ordered phases (for recent reviews, see \cite{neven} and \cite{kezsmarkiPhD}). The driving force of the metal-insulator transition (MIT), and the nature of the order parameters (OP) of the low-$T$ phases is incompletely understood. Our aim is to describe and discuss a number of recent experimental findings and use the results to set up a minimum electronic model of pure BaVS$_3$. With a suitable change of parameters, the same model should serve for the generalized BaVS$_3$ system which includes, along with ideal BaVS$_3$, also sulphur-deficient \cite{defic}, Sr-doped,  and Nb-doped BaVS$_3$, along with the selenide BaVSe$_3$ \cite{bavse}.

The structure of BaVS$_3$ is quasi-one-dimensional: it can be envisaged as a triangular lattice of chains of face-sharing sulphur octahedra. The chains are oriented along the crystallographic $c$-axis which is a $C_3$ axis in the high-$T$ hexagonal phase. V ions sit at the octahedral centres. Within a chain, V--V distances are short (2.81${\AA}$), while the inter-chain separation  is large (6.73${\AA}$). It had been thought that BaVS$_3$ is a quasi-one-dimensional conductor, with a large conductivity anisotropy ($\sigma_c/\sigma_a\gg 1$). However, the first single crystal measurements showed that BaVS$_3$ is an almost isotropic conductor with $\sigma_c/\sigma_a\approx 3-4$ in a wide range of temperature \cite{mihaly2000}. Now we understand that
quasi-one-dimensionality is important in BaVS$_3$ but in a more subtle sense: it leads to the coexistence of two kinds of $d$-electrons.

At ambient pressure, good-quality specimens of stoichioetric BaVS$_3$ undergo three subsequent second order transitions:
a structural (hexagonal-to-orthorombic) transition at $T_{\rm str}=240$K, a metal-insulator transition (MIT) at $T_{\rm MIT}=69$K, and a magnetic ordering transition at $T_X=30$K. The fact that the MIT remains a second order transition in an extended range of pressure \cite{forro00,spingap} shows that the MIT is a symmetry-breaking transition. The nature of symmetry breaking is the lowering of translational symmetry: the unit cell is doubled along the $c$-axis from the high-$T$ two-atomic unit cell to four-atomic unit cell (which may be called tetramers) \cite{inami}.

The succession of three phase transitions means three successive lowerings of the high-temperature symmetry $P6_3mmc{\otimes}R_t$ where $P6_3mmc$ is the space group  and $R_t$ stands for time reversal. Point group symmetry (a $C_3$ rotation) is broken at $T_{\rm str}$, translational symmetry is lowered at $T_{\rm MIT}$ and finally time reversal symmetry is broken at $T_X$ (and at the same time, translational symmetry is further lowered by the creation of a long-period magnetic structure in the $a-b$ plane \cite{magn_str} as well as in the $c$-direction \cite{nakamura}).

Though the crystal lattice changes in at least two of the  three transitions, we may aspire to model all three in terms of a purely electronic model. The $T_{\rm str}=240$K transition can be described as the appearance of orbital polarization in the metallic phase, and the MIT as spin-orbital tetramerization. It is, however, not easy to contrive a lattice fermion model which allows the prediction of the phase diagram of BaVS$_3$.

BaVSe$_3$ is isostructural and isoelectronic with BaVS$_3$ so we may expect that its phase diagram is similar to that of BaVS$_3$ but this is not the case. The structural transition at 300K is followed by ferromagnetic ordering at the Curie temperature $T_C=49$K \cite{bavse}. The system remains metallic at all temperatures.

Since the basic ingredients of an electronic model of BaVSe$_3$ should be the same as for BaVS$_3$, we should require that with tuning the model parameters, the phase diagram of either BaVS$_3$ or BaVSe$_3$ can be reproduced. It transpires that the tetramerization and MIT of BaVS$_3$ require a fine-tuning which is not a generic property of the underlying fermion model. We surmise that in the same sense, a de-tuning of parameters takes place under pressure: the MIT of BaVS$_3$ is suppressed for $p>p_{\rm cr}\approx 2$GPa \cite{forro00}, leaving behind a metal with mysterious magnetic properties \cite{QCP}.

In the spirit of the unified treatment of the classic Mott system
V$_{2-x}$Cr$_x$O$_3$ \cite{Imada,LN}, we extend our interest to the isoelectronically doped systems Ba$_{1-x}$Sr$_x$VS$_3$
\cite{Sr}, and BaV$_{1-x}$Nb$_x$S$_3$ \cite{Nb}. Again, these have  phase diagrams which are substantially different from those of pure BaVS$_3$. Some salient results for Ba$_{1-x}$Sr$_x$VS$_3$ will be described in Sec.~\ref{sec:Sr}.

\section{The $d$-states}

Electronic  structure calculations  show that the Fermi level is pinned in a region with nominal $d$-bands, thus the various forms of BaVS$_3$ do what V$^{4+}$ $d\rightarrow t_{2g}$ electrons can do.  In the trigonal environment of the high-$T$ phase, the $t_{2g}$ level splits into an $A_{1g}$ singlet and an $E$ doublet.

\begin{figure}[h]
\includegraphics[scale=0.7]{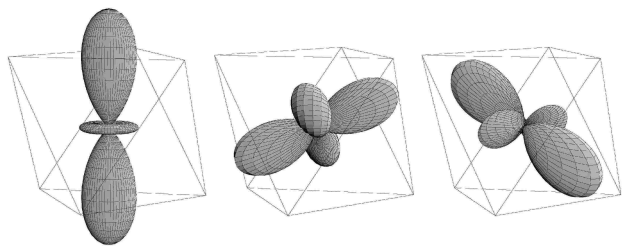}
\caption{Left: The $A_{1g}$ orbitals point towards the next V atom in the octahedral chain. Middle and right: the real basis functions $\phi_a$, $\phi_b$ of the trigonal doublet $E$ point out of the chain direction.}
    \label{fig:orb}
  \end{figure}

The $A_{1g}$ orbital (also called the $z^2$ orbital) is shown in Fig.~\ref{fig:orb} (left). The lobes point across an octahedral face towards the next V atom in the chain. The strong overlap gives rise to the emergence of a 1--2eV wide $A_{1g}$ band.

 The $\phi_a$, $\phi_b$ orbitals of the trigonal doublet $E$
 tilt out of the chain direction (Fig.~\ref{fig:orb}, middle and right), so there is much less direct overlap either along, or between, the chains. Contrasting the small (say, $\sim0.4$eV) $E$-bandwidth with electron--electron interaction energies of the order of several eV, we expect that $E$ electrons are strongly correlated, essentially localized. Indeed optical measurements show that $E$ electrons are localized at all temperatures of interest \cite{optics}.

\begin{figure}[h]
\includegraphics[scale=0.2]{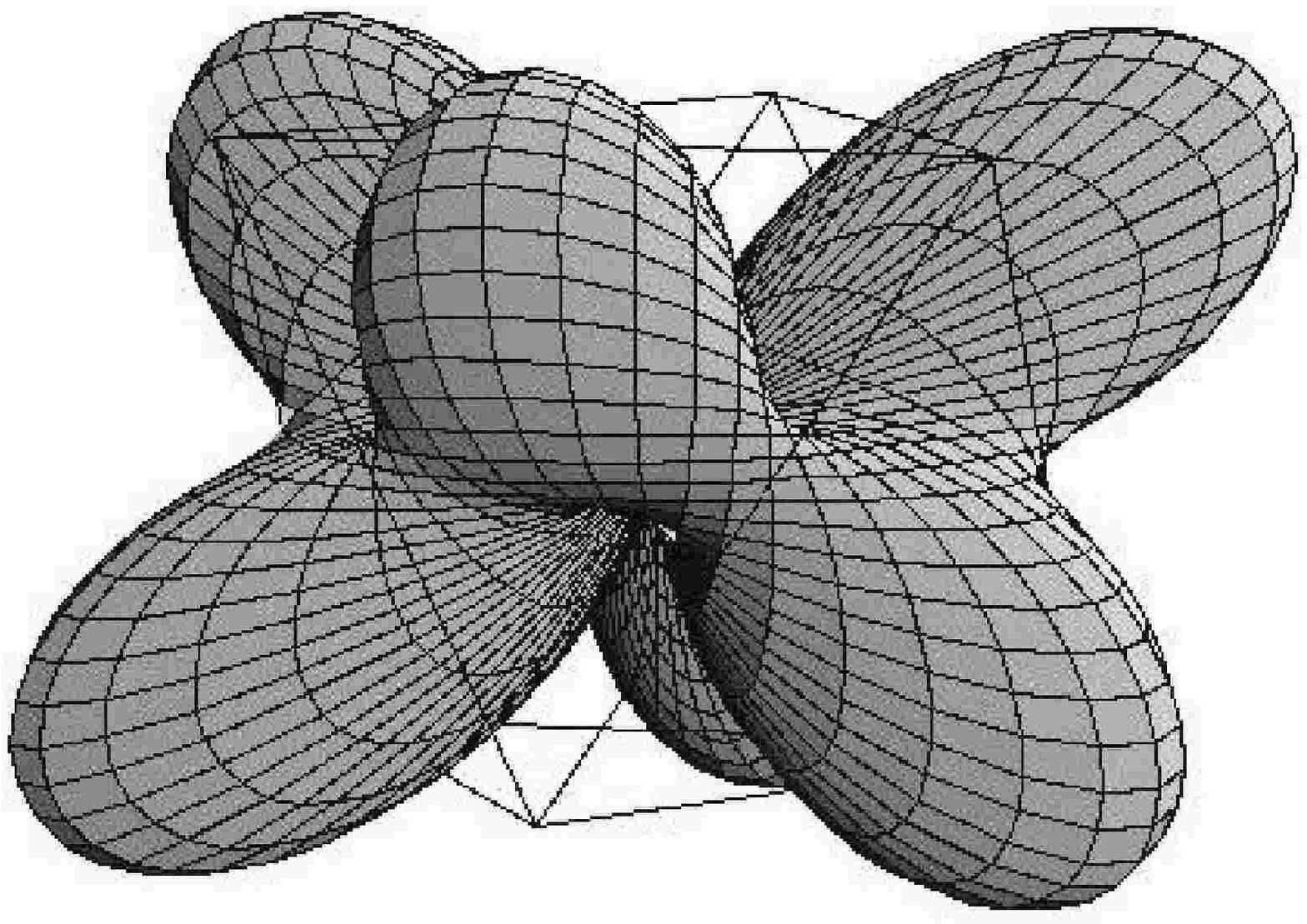}
\includegraphics[scale=0.35]{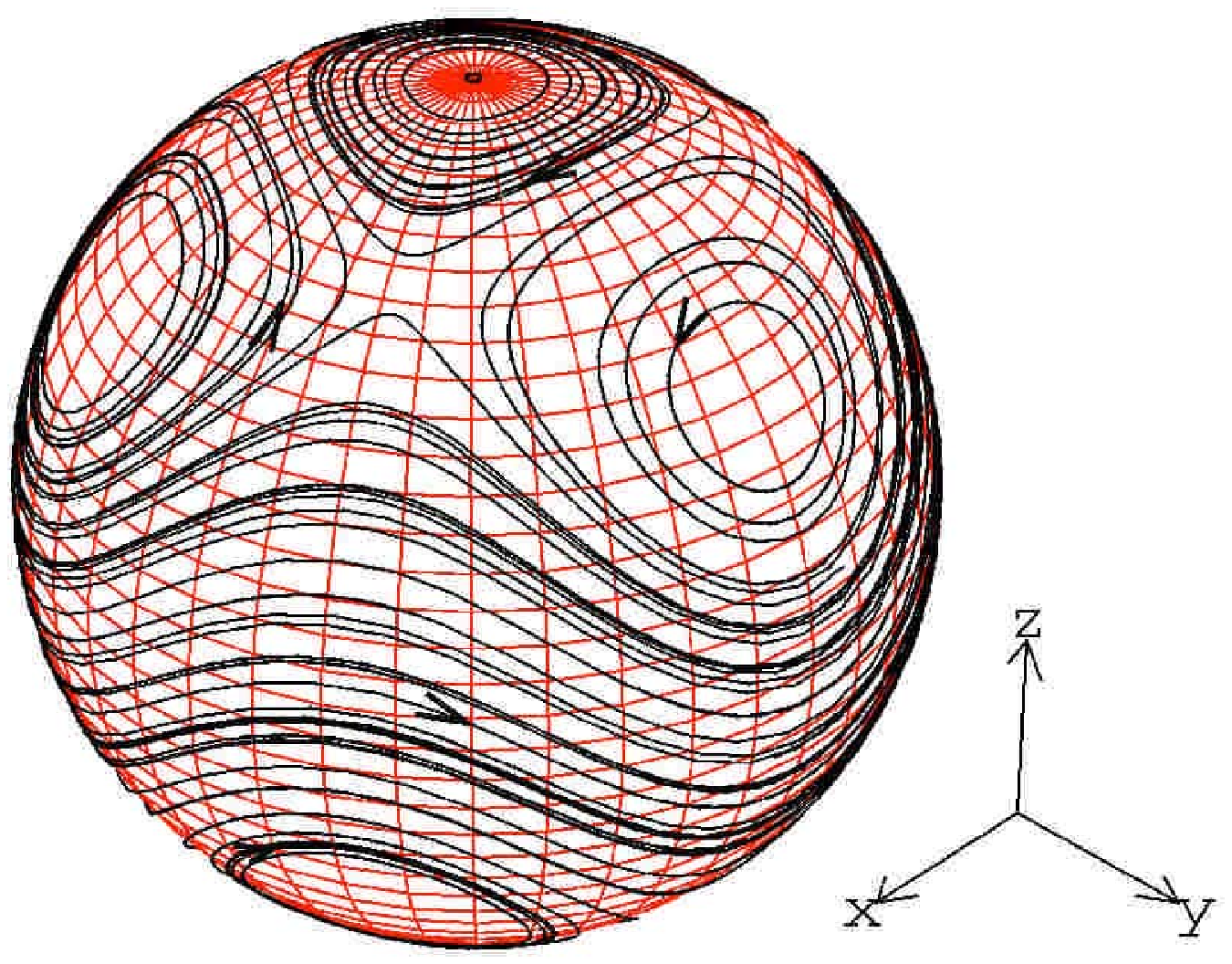}
\caption{The complex combinations $\phi_a\pm {\large I}\phi_b$ of the real trigonal basis functions  $\phi_a$ and $\phi_b$ have the same hexagonal charge cloud (top), but they carry currents (bottom: current flow for $\phi_a + {\large I}\phi_b$. Reversed currents flow in $\phi_a - {\large I}\phi_b$). The current distribution displays both belt-like and eddy currents, illustrating that under trigonal symmetry magnetic dipoles and octupoles mix.}\label{fig:current}
 \end{figure}

Which localized degrees of freedom may the $E$ electrons carry? The real basis functions $\phi_a$, $\phi_b$ shown in Fig.~\ref{fig:orb} (right) carry quadrupolar moments. However, this is not the only possibility. Within the doublet, we are free to choose their complex linear combinations $\Phi_{\pm} = (1/\sqrt{2}) (\phi_a \pm I\phi_b)$ as a basis. $\Phi_{+}$ and $\Phi_{-}$ have the same electron density (Fig.~\ref{fig:current}, top), but they carry non-zero intrashell currents (Fig.~\ref{fig:current}, bottom), whereby $\Phi_{+}$ is the time-reversed of $\Phi_{-}$.

Quadrupolar moments are quenched in the complex states, but we have a kind of an orbital angular momentum. Its nature can be guessed from the current flow lines shown in Fig.~\ref{fig:current}.
Part of the current encircles the core: it corresponds to net angular momentum. In fact  $\langle\Phi_{\pm}|L^z|\Phi_{\pm}\rangle=\pm 1$, where $L^z$ is the component of the angular momentum along the chain direction. Let us, however, note the presence of eddy currents in
Fig.~\ref{fig:current}. Such eddies are familiar from octupolar states
\cite{fkr}, and here the illustrate the statement that magnetic dipoles and octupoles are mixed under trigonal symmetry.

\begin{figure}[h]
\includegraphics[scale=0.7]{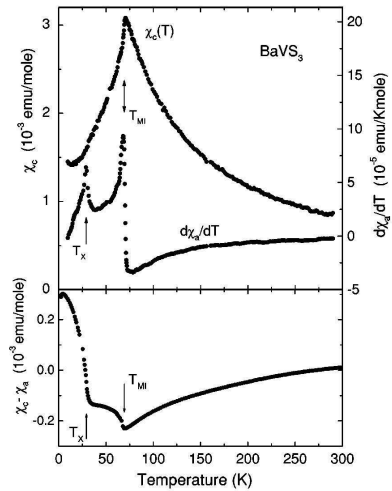}
\caption{Temperature dependence of the magnetic susceptibility, and its anisotropy (after \protect\cite{mihaly2000}).}
\label{fig:susc}
\end{figure}

To summarize, thinking of purely local (on-site) orbital order, $E$ electrons may support either quadrupolar (real orbital) order, or complex orbital order \cite{CO} which is a mixture of Ising-like dipolar order with octupolar order. The $A_{1g}$ electrons have no local orbital degree of freedom. In addition, both $E$ and $A_{1g}$ electrons have spins which may order either independently of, or simultaneously with the orbital degrees of freedom. Though there may be transitions which are predominantly orbital, and others which are predominantly spin-ordering phenomena, on general grounds we should expect that spin and orbital order mutually induce each other, subject to symmetry restrictions \cite{anis}. The relativistic spin--orbit coupling is not negligible for V ions. The anisotropy of the spin susceptibility $\chi_c-\chi_a$  is in the range $\pm 25${\%}, and shows sharp anomalies at the MIT, and the magnetic ordering transition (Fig.~\ref{fig:susc}, bottom).

In addition to on-site order parameters, there are local order parameters involving a pair of sites (e.g. those characterizing the internal structure of the crystallographic unit cell which contains two V atoms), or four consecutive sites along a chain (this is important for
the tetramerized insulating phase in the temperature range
$T_X=30{\rm K}<T<T_{\rm MIT}=69{\rm K}$). For the metallic orthorombic phase at $T_{\rm MIT}=69{\rm K}<T<T_{\rm str}=240{\rm K}$ an itinerant order parameter like the orbital polarization of the $E$ subband is more adequate. The number of possible ordering schemes is considerable, and different members of the family of BaVS$_3$/BaVSe$_3$-based materials belong to different realizations of the phase diagram.

\section{Experimental band structure: ARPES}\label{sec:arpes}

In principle, the formal $d$-electron count V$^{4+}\leadsto 3d^1$ could be satisfied with $A_{1g}$ electrons alone ($n(A_{1g})=1$ where $n$ is number/V-atom), or $E$-electrons alone $n(E)=1$, or any ratio $n(A_{1g})/n(E)$ requiring that $n(A_{1g})+n(E)=1$. However, it can be argued that, on the one hand, $E$ electrons must be present in BaVS$_3$ and, on the other hand, $E$-electrons alone could not account for the observed behavior. The hexagonal-to-orthorombic transition at $T_{\rm str}=250$K makes energetic sense only if it serves to lift the degeneracy of the $E$ states. The large Curie-like susceptibility at $T>T_{\rm MIT}$ (Fig.~\ref{fig:susc}, top) can be interpreted as the spin susceptibility of localized $E$-electrons if their density is about $n(E)\sim 1/2$.  On the other hand, $E$-electrons only could not account for the conductivity of the hexagonal metallic phase, so the Fermi level must lie within the $A_{1g}$ band.

LDA calculations  arrive at the estimate $n_{\rm LDA}(A_{1g})\approx 0.72$  \cite{mattheiss,Georges}. The corresponding concentration $n_{\rm LDA}(E)\approx 0.28$ of localized electrons could not be reconciled with the observed susceptibility. The difficulty is resolved by DMFT calculations which show that intra- and inter-site correlation effects cause a redistribution of electrons over the subbands so that in the low-$T$ phases $n(A_{1g})\approx n(E) \approx 1/2$ is realized. Such a magic value of the $n(A_{1g})/n(E)$ ratio allows the arising of short-period ordered states, e.g., tetramerized states where the four-atom unit cell may be either  $AE_aAE_b$ (as suggested in Ref.~\cite{RXS}), or $AEEA$ (which we discuss as an alternative). In any case, the magic value $n(A_{1g})/n(E)\approx 1$ seems to be a precondition for opening a gap in the continuous MIT. In other words, correlation acts to create an electronic structure in which a low-$T$ insulating phase is possible. In this sense  we may consider BaVS$_3$  strongly correlated  though it does not have an insulating phase without some kind of symmetry breaking.

  \begin{figure}[h]
\includegraphics[scale=0.8]{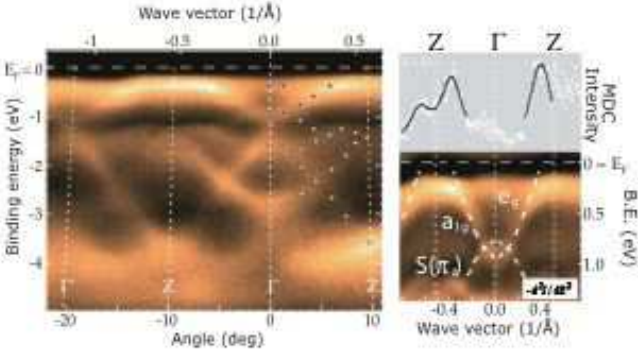}
\caption{(Left) ARPES intensity map of BaVS$_3$ measured along the $\Gamma$-Z direction at $T = 40$K. The intensity scales from black to white. Circles show the
positions of peaks in corresponding energy distribution curves
(photoelectron intensity vs. binding energy). The white circles follow
predominantly S(3p) bands and the black ones V(3d) bands.  (Right below). Second derivative of a detail of the map on the left, and in the proximity of the Fermi level. The image shows both V(3d) bands: the weakly dispersive $E$ bands, and the dispersive $A_{1g}$ band that is hybridized with a S($\pi_z$) band. (Right above) The momentum distribution curve (photoelectron intensity
vs. wave vector) taken at the Fermi energy with 50 meV integration shows the crossings of the $A_{1g}$ band at around 3/4 of $\Gamma$-Z. No crossing is easily detectable for the $E$ part.
(after\protect\cite{ARPES}). }
    \label{fig:arpes}
  \end{figure}

It is important to have direct experimental information about the relative position of the $A_{1g}$ and $E$ bands. An
angle-resolved photoemission spectroscopy (ARPES) experiment was
carried out \cite{ARPES} which reveals the nature of the band structure for $k\parallel{c^*}$. Representative results are shown in Fig.~\ref{fig:arpes}. The essential result is that the Fermi level is pinned in an energy range with high density of state where a dispersive $d$-band (the $A_{1g}$ band) crosses a set of levels whose dispersion could not be directly measured but we may estimate that the associated band-width is at most 0.3eV. This fits our idea of $E$-like states: DMFT gives very narrow effective bands \cite{Georges} while IR optics shows that $E$-states are essentially localized \cite{optics}. The ARPES experiment gave information about the $k\parallel{c^*}$ dispersion only, so we could not directly measure
$n(A_{1g})/n(E)$ which requires a Brillouin zone average. Nevertheless, we can assert that neither $n(A_{1g})$ nor $n(E)$ is small and that $n(A_{1g})/n(E)\approx 1$ is at least compatible with the experiment.

The finding that essentially the same number of wide-band and narrow-band $d$-electrons coexist at energies near the Fermi level,
allows to speculate about the nature of the relevant fermion models of BaVS$_3$ (meaning both pure BaVS$_3$ and the related systems).
It has to be a two-band model, a kind of a $d$-band Anderson lattice, with the $A_{1g}$-states playing the role of the wide band, and the $E$-like states the role of the strongly correlated narrow band.
The next question is that whether the $E$-level should be considered orbitally degenerate. Strictly speaking,  the orthorombic distortion setting in at $T_{\rm str}=250$K lifts the degeneracy of the real $E$-states $\phi_a$ and $\phi_b$, and one may be tempted to omit one of the $E$-orbitals. However, this would prohibit the creation of the orbital-momentum-carrying complex states (Fig.~\ref{fig:current}), and thus would leave us without explanation of the susceptibility anisotropy (Fig,~\ref{fig:susc}) which we ascribe to the spin-orbit coupling. Therefore we suggest that the effective hamiltonian should include the orbital degeneracy of the $E$-electrons and explain the orbital polarization as an interaction effect. Schematically, the effective  hamiltonian (written, for the sake of simplicity, in one-dimensional form), is
\begin{eqnarray}
H = H_{\rm PAM} + H_{\rm Hund} + H_{\rm int}
\end{eqnarray}
where $H_{\rm PAM}$ is the periodic Anderson model (PAM) part discussed above, $H_{\rm Hund}$ is the Hund coupling, and $H_{\rm int}$ contains additional interactions. The PAM part is
\begin{eqnarray}
H_{\rm PAM} & = & -t_A \sum_{j\sigma}
(a_{j\sigma}^{\dagger}a_{j+1\sigma} +H.c.) + \epsilon_E \sum_j \sum_{\alpha\sigma} n_{E,j,\alpha,\sigma} \nonumber \\
& & +U \sum_j \sum_{\gamma\sigma} \sum_{\gamma'\sigma'} n_{j,\gamma,\sigma}n_{j,\gamma',\sigma'}
\end{eqnarray}
where $a_{j\sigma}^{\dagger}$ creates an $A_{1g}$ electron at site $j$
with spin $\sigma$, $e_{j\alpha\sigma}^{\dagger}$ the same for an $E$ electron with orbital index $\alpha=a,b$, the $n$'s are corresponding occupation numbers, and in the Hubbard term $\gamma$ incorporates both $A_{1g}$ and $E$ electrons. We neglected $E$-hopping. We expect that double occupation by $E$-electrons is suppressed and the Hund coupling counts only if an $E$-electron  and an $A$-electron share a site:
$H_{\rm Hund} = -J_{\rm H}\sum_j {\bf S}_{j,A}{\cdot} {\bf S}_{j,E}$.
Double exchange may well explain the itinerant ferromagnetism of
BaVSe$_3$ \cite{bavse} and BaVS$_{3-x}$ \cite{defic}. It may play a delicate role by inducing ferromagnetic correlations in an overall
non-ferromagnetic state in pure BaVS$_3$.

We envisage that locking in to the magic ratio \newline
$n(A_{1g})/n(E)\approx 1$ ($n(A_{1g})+n(E)=1$ being fixed) emerges in only certain phases of $H$. In any case, it has to be an intersite interaction effect. In the strong coupling limit $H_{\rm PAM}+H_{\rm Hund}$ generates a hierarchy of effective interactions, including multi-site interactions. These should be combined with intersite Coulomb processes which we collect in $H_{\rm int}$.

In particular we expect that an effective interaction for nearest neighbour $EE$ pairs appears, with the general structure that the orbital interaction ($I_s$ or $I_t$, resp.) is different in the spin-singlet ($s$) and spin-triplet ($t$) sectors
\begin{eqnarray}
H_{EE} & = & J_s\sum_j \left( \frac{n_j^E n_{j+1}^E}{4}-{\bf S}_{j,E}{\cdot}{\bf S}_{j+1,E} \right) I_s(\tau_{j,E},\tau_{j+1,E})\\
 & & + J_t\sum_j
\left(\frac{3}{4}n_j^E n_{j+1}^E+{\bf S}_{j,E}{\cdot}{\bf S}_{j+1,E})\right) I_t(\tau_{j,E},\tau_{j+1,E})\nonumber
\end{eqnarray}
If $J_s$ is stronger than $J_t$, the system will contain a number of
$EE$ singlet pairs. This may explain why a spin gap opens at the MIT \cite{spingap}. The density of $EE$ pairs and their spatial distribution will be decided by effective multi-site interactions which in turn reflect the polarizability of the $A_{1g}$ band at the particular filling. All in all, promotion of electrons from the $A_{1g}$ band to localized $E$ levels is  favored by two sources of energy gain: $E-E$ singlet binding, and the opening of a gap in the band.

It is clear from the susceptibility curve (Fig.~\ref{fig:susc}) that as
$T_X=30$K is approached from above, a substantial number of $E$-spins is not in single pairs yet. These residual spins order magnetically, assuming a thee-dimensional long-period structure but we may suspect that this complicated structure is picked by weak residual interactions and it is the least robust feature of BaVS$_3$.

\section{Magnetic properties under pressure}

The properties of stochiometric BaVS$_3$ are sensitive to pressure. The MIT is gradually suppressed until it vanishes at the critical pressure $p_{\rm cr}\approx 2$GPa \cite{forro00} (see also the $x=0$ curve in the inset of Fig.~\ref{fig:gauzzi1}).  Unfortunately no high-pressure susceptibility measurements have been carried out yet so we have to infer the magnetic character of high-$p$ (either insulating or metallic) BaVS$_3$ from transport measurements.

At ambient pressure AFM order sets is at $T_X=30$K. We have no own susceptibility data under pressure, but we learned from a most interesting private communication by H. Nakamura and T. Kobayashi that $T_X(p)$ remains essentially unchanged in a finite pressure interval. However, there are no known susceptibility results which would give direct evidence whether $T_X(p)$ reaches, or crosses, $T_{\rm MIT}(p)$.

\begin{figure}[ht]
 \includegraphics[scale=0.45]{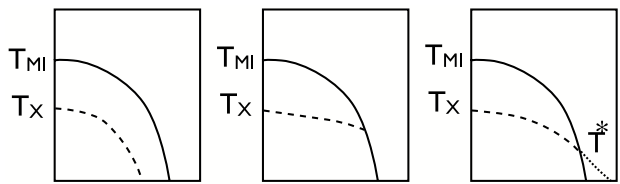}
\caption{Three principal possibilities for the high-$p$ part of the phase diagram: $T_X\to 0$ before $x \to x_{\rm cr}\approx 20$GPa (top); the low-$T$ insulator is always magnetic though the high-$p$ metal does not order magnetically (middle); or there is a high-$p$ magnetic phase but the ordering at $T_X^*$ differs in some respects from the ordering of the insulator at $T_X$ (bottom).}\label{options}\label{fig:tript}
 \end{figure}

We discuss the three alternatives shown in Fig.~\ref{fig:tript}:
\begin{enumerate}
\item{At high pressures, the non-magnetic tetramerized state becomes the ground state (Fig.~\ref{fig:tript}, top). $T_{\rm MIT}$ is always distinct from $T_X$,  the MIT retains the same character at all pressures.}
\item{$T_X(p)$ meets $T_{\rm MIT}(p)$ somewhere below $p=$2GPa at a multicritical point (Fig.~\ref{fig:tript}, middle). The low-$T$ insulator is always magnetic, or in other words, the high-$p$ part of the MIT boundary belongs to a metal-to-magnetic-insulator transition. There should be a change of character of the MIT at the point where
phase boundaries cross. }
\item{There is low-$T$ magnetic order in the high-$p$ metal (Fig.~\ref{fig:tript}, bottom). It need not be the same magnetic order as observed at ambient pressures; in fact, since the low-$p$ magnetic phase grew out of a pre-existing tetramerized background which is not present at $p>p_{\rm cr}$, we have reason to expect that the two magnetic orders are different. To emphasize this, we denoted the high-$p$ magnetic transition temperature with $T_X^*$. }
\end{enumerate}

In what follows, we present magnetoresistivity data which prove that $T_X(p)$  gets near $T_{\rm MIT}(p)$ (so at least excluding Option i)). There is an additional finding indicating that in fact Option iii) belongs to the true situation, i.e., there is an ordered phase at $T<T_X^*$ in a pressure interval above the critical pressure. Finally, we describe
Sr-doped BaVS$_3$ which can be viewed as BaVS$_3$ under chemical pressure.

\subsection{High-pressure magnetoresistivity experiments on BaVS$_3$}

\begin{figure}[ht]
 \includegraphics[scale=0.4]{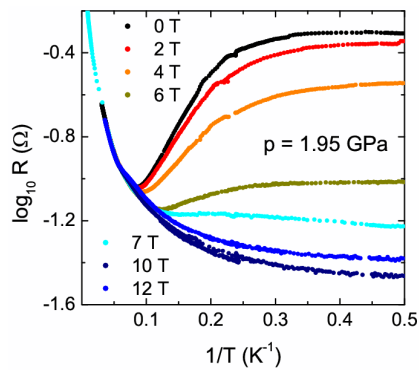}
 \caption{At marginally subcritical pressures, transition into the metallic state can be induced by an applied magnetic field. The critical field is about 7T. (after \protect\cite{field}.)} \label{fig:fieldinduced2}
 \end{figure}

In the high-pressure experiments described in Ref.~\cite{field} we
found that at carefully chosen pressures $T_{\rm MIT}$ can become as low as 7-10K. With $k_{\rm B}T_{\rm MIT}$ of the same order as Zeeman splittings in laboratory fields, it is reasonable to ask whether the MIT can be induced by magnetic field. Such is indeed the case. Fig.~\ref{fig:fieldinduced2} shows an example when in the absence of field, the sample is insulating, and increasing the magnetic field changes the character of the $T$-dependence from non-monotonic (insulating at low $T$) to monotically decreasing, i.e., metallic at all $T$.

Suppression of the insulating phase by field is plausible if the insulator has antiferromagnetic correlations. This would be true also of the
tetramerized spin-gapped phase but a detailed argument \cite{field} shows that the field-induced reduction of $T_{\rm MIT}$ is different from that seen at lower pressures \cite{spingap}. Thus we conclude that we observe the suppression of a magnetically ordered phase.

\begin{figure}[h]
 \includegraphics[scale=0.8]{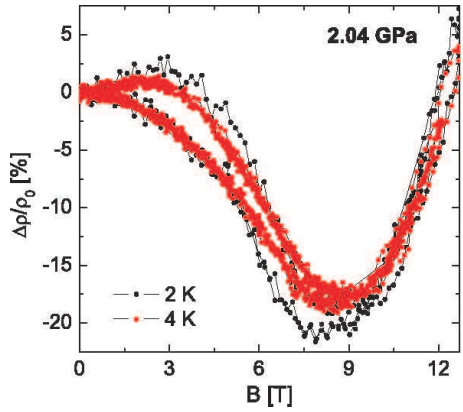}
\caption{The presence of a hysteresis cycle in the
magnetoresistivity proves the existence of a symmetry-breaking state
in the high-pressure metallic phase (after
\protect\cite{QCP,neven}).}\label{fig:hyst}
 \end{figure}

Next, we ask if the high-$p$ metal undergoes an ordering transition at
low $T$. At a pressure slightly in excess of $p_{\rm cr}$ we observed a hysteresis loop in the field dependence of the resistivity (Fig.~\ref{fig:hyst}). Such a history dependence of a physical quantity indicates an underlying ordering phenomenon. We conclude that (in some pressure range at least) high-$p$ BaVS$_3$ has an ordering transition, most likely a magnetic ordering transition.

The existence of a  $T_X^*$ transition is plausible on general grounds.
Even though the metallic phase avoided the structural instability of tetramerization, it may still break time reversal invariance.  For instance, BaVSe$_3$ does not tetramerize (and thus remains metallic)
but it becomes ferromagnetic at $T_C=49$K \cite{bavse}. We may ask whether high-$p$ BaVS$_3$ is simply a ferromagnet. This is unlikely.
The suspected order is in the same $(p,T)$ regime where the resistivity shows non-Fermi liquid (NFL) behavior with $\Delta\rho\propto T^n$, with $n<2$ \cite{QCP}. NFL behavior is not
routinely associated with itinerant ferromagnetism. The low-$T$ ordering of the high pressure metallic phase must be of complex nature.

\subsection{Sr-doped BaVS$_3$}
\label{sec:Sr}

Sr is isovalent with Ba, but much smaller. Thus replacing part of Ba atoms with Sr contracts the lattice due to "chemical pressure". A unified view of the effects of hydrostatic and chemical pressure proved fruitful in the study of V$_{2-x}$Cr$_x$O$_3$ \cite{Imada,LN}.
It is of obvious interest that magnetic measurements are possible for ambient-pressure Ba$_{1-x}$Sr$_x$VS$_3$ which we may regard as a surrogate high-pressure BaVS$_3$.

 \begin{figure}[h]
 \includegraphics[scale=0.27]{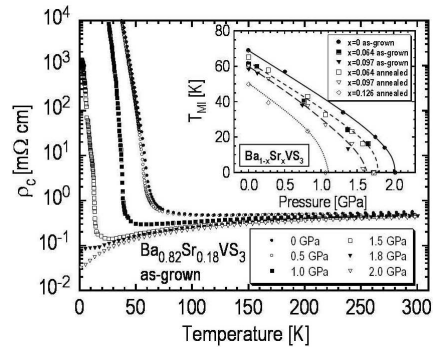}
\caption{Resistivity vs $T$ plots for Ba$_{0.82}$Sr$_{0.18}$VS$_3$ at selected pressures. Inset: metal-insulator transition  phase boundaries in the $p$-$T$ plane for selected compositions (after \protect\cite{Sr}).}\label{fig:gauzzi1}
 \end{figure}

\begin{figure}[h]
 \includegraphics[scale=0.37]{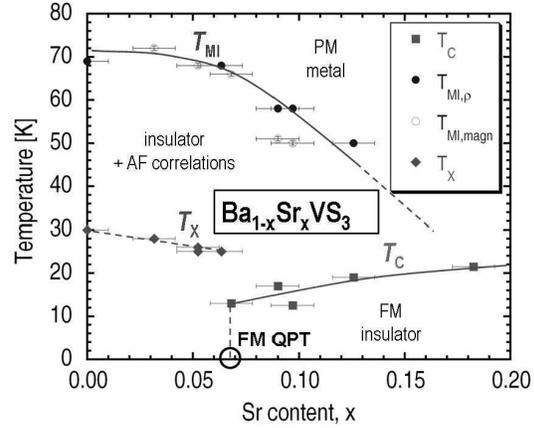}
\caption{Ambient pressure phase diagram of Sr-doped BaVS$_3$ in the $T$--$x$  plane on the basis of resistivity and susceptibility measurements (after \protect\cite{Sr})}\label{fig:gauzzi3}
 \end{figure}

Fig.~\ref{fig:gauzzi1} shows the plots of resistivity vs $T$ for various Sr-doped samples under pressure. It appears that as far as the MIT is concerned, chemical pressure from Sr content, and externally applied pressure can be added together. However, this holds only conditionally if the low-$T$ magnetic phase is considered. Susceptibility measurements \cite{Sr} show that at low doping the character of the $T_X$ transition is approximately preserved, so we may assume a long-period antiferromagnetic ground state. However, at $x_{\rm cr}\sim 0.07$ there is a quantum phase transition: a sudden switch to a ferromagnetic ground state. In contrast to other examples of ferromagnetism in the BaVS$_3$ system, this time we have a FM insulator. A preliminary ambient pressure phase diagram based on samples with $x<0.18$ is shown in Fig.~\ref{fig:gauzzi3}.

{\bf Acknowledgments}. The authors acknowledge support by the Hungarian National Grants OTKA K62280, K62441, TS049881, and Bolyai 00239/04.

\newpage

\end{document}